\documentclass[preprint,nofootinbib,tightenlines,showpacs]{revtex4}
\usepackage{amssymb}
\usepackage{amsmath}
\usepackage{lscape}
\usepackage{epsfig}

\def\journal#1, #2, 1#3#4#5, #6{
    {\sl #1~}{\bf #2} (1#3#4#5) #6}

\def\beq{\begin{equation}}
\def\eeq{\end{equation}}
\def\ba{\begin{eqnarray}}
\def\ea{\end{eqnarray}}

\def\l{\lambda}
\def\o{\omega}

\def\r{\rho}
\def\p{\partial}

\newcommand{\h}{Hamiltonian}

\newcommand{\baa}{\begin{eqnarray}}

\begin{document}
\preprint{CERN-PH-TH/2007-253}

\title{Homolumo Gap and Matrix Model}
\author{I. Andri\'c}
\author{L. Jonke}
\author{D. Jurman}
\affiliation {Theoretical Physics Division, Rudjer Bo\v skovi\'c Institute, P.O. Box 180, 10002 Zagreb, Croatia}
\author{H. B. Nielsen}
\affiliation{The Niels Bohr Institute, Copenhagen DK 2100, Denmark}
\altaffiliation{On leave of absence to CERN until 31 May, 2008.}
\begin{abstract}
We discuss a dynamical matrix model by which probability distribution is associated with Gaussian ensembles from random matrix theory.
We interpret the matrix $M$ as a Hamiltonian representing interaction of a bosonic system with a single fermion. We show that a system of second-quantized fermions influences the ground state of the whole system by producing a  gap between the highest occupied eigenvalue and the lowest unoccupied eigenvalue.
 \end{abstract}

\maketitle

\section{Introduction}

Random matrices \cite{mehta} first studied by Wigner \cite{wigner} have applications in many branches of (many-body) physics, e.g., nuclear and molecular physics. The matrix elements  are considered statistical variables and take random values. On the other hand, one can consider a dynamical matrix model in which the matrix elements are dynamical variables so that the whole matrix has become a mechanical system, which may even be quantized and corresponding quantum field theory may be formulated. Such dynamical matrices have found use in e.g. description of disordered system and in string theory \cite{Alexandrov}.

In the present work we treat a dynamical matrix model by which probability distribution in the ground state is associated with Gaussian ensembles from random matrix theory.
We interpret the matrix $M$ as a \h \ representing interaction of a bosonic system, such as for example the nuclei in a molecule,  with a single fermion, e.g. a single electron in a molecule. In particular, an eigenvalue of the matrix correspond to an eigenenergy of a fermion.
Without going into so much detail of the single fermion interaction, we consider a backreaction from the system of the second quantized  fermions.
A characteristic effect of such a backreaction is to drive the matrix in the direction of lowering the energy of the eigenstates occupied by fermions. In the case of a molecule this backreaction is the force from the electrons pushing on the nuclei seeking to drive them into such a position so as to lower the filled single electronic orbit energy eigenvalues. Since a similar push to lower the unoccupied levels is absent, or even has the opposite sign, as will be argued later, hereby can easily arose a gap between the highest occupied eigenvalue (homo, where mo stands for molecular orbit) and the lowest unoccupied eigenvalue (lumo). The gap is what is called the homolumo gap \cite{jt}. The interest in the homolumo gap steams from the fact that the details of the model seem very unimportant. You should get it whenever you have a system of "bosonic" variables interacting with the fermions, provided that the "bosonic" variables are sufficiently soft as to yield to the pressure from the fermions.

The purpose of this paper is to evaluate how distribution of eigenvalues of the matrix adjust to produce a homolumo gap in quite general setting.
First, we formulate field theory corresponding to the large N limit of the random ($N\times N$) matrix. Then we introduce the dynamical matrix model by which probability distribution in the ground state coincides with the large N limit of the 'free' Gaussian ensemble from the random matrix theory. The notion of the dynamical model enables us to model the interaction potential which arose as a consequence of the backreaction of the second quantized fermions. Finally, exploring the effect of adding this interaction potential to the 'free' matrix model, we obtain characteristic homolumo gap.

\section{Field theory formulation of the random matrix model}
The field theory we discuss is defined by the following functional integral
\begin{equation}\label{zz}
Z[\r]= \int {\cal D} \r \; \exp\left\{(\l-1)\int dx\r(x)\ln \r(x)+\l \int dx dy \r(x)\ln (x-y)\r(y)-\int dx \r(x) P(x)\right\},
\end{equation}
for some (polynomial) function $P(x)$ and a dimensionless parameter $\l$.
The integral (\ref{zz}) corresponds to the large $N$ limit of the matrix integral from random matrix theory
\begin{equation}\label{ztr}
Z=\int dM e^{-TrP(M)},
\end{equation}
where $P(M)$ is a polynomial in $N\times N$ matrix $M$.
In terms of eigenvalues $x_i$ of the matrix $M$, the matrix integral (\ref{ztr}) can be expressed as
\begin{equation}\label{rmt}
Z=\int \prod_{i} dx_i \prod_{i< j}(x_i-x_j)^{2
\l} e^{-\sum_i P(x_i)}.
\end{equation}
Here, the parameter $\l=1/2,1,2$ for real-symmetric, hermitean and quaternionic-real matrix $M$, respectively. 
The origin of the first two terms in (\ref{zz}) is  the new invariant measure resulting  from changing the variables from $x_i$ to $\r(x)$, see Refs.\cite{dy,meas}: 
\begin{equation}\label{po}
\prod_{i< j}(x_i-x_j)^{2\l} \to \exp\left\{\l\int dx \r(x)\ln\r(x)+\l\int dx dy\r(x)\ln(x-y)\r(y)\right\}.  
\end{equation}
\begin{equation}\label{mj}
\int \prod_{i}^N dx_i \to \int {\cal D}\r\exp\left\{-\int dx \r(x)\ln\r(x)\right\}.
\end{equation}
While in (\ref{rmt}), the statistical variables are the eigenvalues $x_i$, in the large $N$ limit one introduces the density field $\r(x)$
which then by itself becomes statistical variable. 
In parallel with Ref.\cite{th}, we interpret (\ref{zz}) as a quantum field theory describing statistical random matrix model (\ref{ztr}).
By varying (\ref{zz}) with respect to $\r$ and taking derivative with respect to 
$x$ we find that the most probable configuration satisfies following equation
\begin{equation}\label{BPS}
(\l-1)\frac{\p_x \r}{\r}-2\l \pi \r^{H}=\p_x P,
\end{equation}
where $f^H$ denotes Hilbert transform of the function $f$.
From (\ref{BPS}) we  obtain the Riccati differential equation for the new field $W=\r^{H}+i\r$ in analogy with \cite{jhep}:
\begin{equation}\label{rW}
(\l-1)\p_x W-\l \pi W^2-W\p_x P =(\r\p_x P)^{H}-\r^{H}\p_x P
\end{equation}
This equation corresponds to the usual Riccati equation for the resolvent function in the random matrix theory except that here the first term vanishes {\it exactly} for the hermitean matrix model ($\l=1$). The difference comes from the additional term in measure (\ref{mj}).

Before introducing the matrix model which describes interaction of fermions and bosons we need the notion of dynamical system to be able to define and model the potential, and also, to be able to interpret the integrand in (\ref{zz}) as a probability density functional in the ground state of this dynamical model. This  requirement is satisfied by the following Hamiltonian (see \cite{jhep} for details): 
\begin{equation}\label{dyn}
H=\frac{1}{2}\int dx \r(x) A^{\dagger}(x)A(x) +E_0,
\end{equation} 
where $A(x)$ is given by
\begin{equation}
A(x)=\p_x \pi (x) + i \p_x\frac{\delta \ln \Phi}{\delta \r(x)},
\end{equation}
for $\Phi$ being the ground-state functional (i.e., the square root of the integrand in (\ref{zz})) and $\pi(x)$ the canonical momentum  satisfying $[\p_x\pi(x),\r(y)]=-i\p_x\delta(x-y)$.
Now, we interpret Eq.(\ref{BPS}) as a BPS equation of motion. 
Let us consider a simple example of the 'free' Gaussian ensemble defined by (\ref{rmt}) with $P(x)=x^2$. The operator $A(x)$ of the corresponding dynamical model is 
\begin{equation}
A(x)=\p_x\pi(x)+i\left(\frac{\l-1}{2}\frac{\p_x\r(x)}{\r(x)}-\l\pi\r^H(x)-x\right),
\end{equation}
and the \h \ can be written as 
\begin{eqnarray}\label{cll}
{ H}=\frac{1}{2}\int dx \r(x)(\p_x\pi(x))^2+\frac{1}{2}\int dx \r(x)
\left(\frac{\l-1}{2}\frac{\p_x\r(x)}{\r(x)}-\l\pi\r^H(x)-x\right)^2+E_0,
\end{eqnarray}
The second term is called effective potential $V_{\rm eff}$ and represents the physical potential for the dynamical model in question.
The ground-state functional satisfying  $A(x)\Phi=0$ is 
\begin{equation}
\Phi=\exp\left\{\frac{(\l-1)}{2}\int dx\r(x)\ln \r(x)+\frac{\l}{2} \int dx dy \r(x)\ln (x-y)\r(y)-\frac{1}{2}\int dx \r(x)x^2\right\},
\end{equation}
giving the integrand of the (\ref{zz}) as a probability density functional in the ground state.
The semiclassical solution is given by the solution of the  equation (\ref{BPS}), which in  $\l=1$ case  reads $-\pi  \r^H(x)= x$ and gives   
 the usual Wigner semicircle law for the distribution of eigenvalues for the hermitean Gaussian ensemble. In the rest of the paper we restrict ourselves to the $\l=1$ case.

\section{Interaction with fermions}

It is simply the idea of the interaction of the already described dynamical matrix model with the system of fermions that we postulate that dynamical matrix  itself is the \h \ for a single fermion. Then on top of that we second-quantize the fermions so that it becomes possible to have filled or empty all the (basis) states for this matrix. If we especially seek the ground state of the whole system - dynamical matrix plus fermions - we shall be interested in the case when the eigenstates of the dynamical matrix are filled below a certain value $x_{F}$, the fermisurface (value), while the ones above $x_F$ are empty. In case we could ignore that this fermisurface (energy) depends on the (dynamical) state of the  matrix  the interaction with the fermions simply gives an extra potential 
\begin{equation}
V_{\rm fermion}(x)=x\theta(x_{F}-x).
\end{equation}

At this point we introduce  the zero-point energy for fermion modes.
If we want to write a hermitean \h \ contribution from a single particle state  as a bilinear expression in the annihilation $a$ and creation $a^{\dagger}$ operators and symmetrized in taking the product we must take the commutator expression
\begin{equation}
H=\frac{\o}{2}[a^{\dagger},a]=\o a^{\dagger}a-\frac{\o}{2}.
\end{equation}
The extra term $-\o/2$ compared to the pure number operator 
term, $\o N=\o a^{\dagger}a$, is analogous to the zero-point energy term $\o/2$ for a boson model 
\begin{equation}
H=\frac{\o}{2}\{a^{\dagger},a\}=\o a^{\dagger}a+\frac{\o}{2}.
\end{equation}
With this analogy between bosons and fermions in mind we can claim that there is a zero-point energy $-\o/2$ for fermion mode meaning that the energy of an empty level is indeed $-\o/2$, rather than the naive zero. Then of course the energy of the filled level should  be  $\o/2$. 


The system which we consider of a dynamical matrix $M$ interacting with a system of second quantized fermions described by annihilation $a_i$ and creation $a_j^{\dagger}$ operators denoted by the same indices $\{i,j\}$ as the columns and rows in the dynamical matrix $M$. This means - with the convention of inclusion of the zero-point energy - that the energy term for the interaction of the dynamical matrix $M$ with the fermions, and that includes actually all energy of the fermions, becomes
\begin{equation}\label{neki}
H_{\rm int.\;fermion}=\frac{1}{2}[(a_1^{\dagger},a_2^{\dagger},\hdots,  a_N^{\dagger}),M(t)\begin{pmatrix}a_1 \\ a_2 \\ \vdots \\ a_N\end{pmatrix}].
\end{equation}
If we let $a_i$ and $a_i^{\dagger}$ be annihilation and creation operators for the eigenstate of matrix $M$ instead of for the original basis vectors,  \h \ (\ref{neki}) would reduce to 
\begin{equation}\label{pma}
H_{\rm int.\;fermion}=\sum_i\frac{1}{2}x_i\left[a_i^{\dagger},a_i\right]=\sum_i x_i(a_i^{\dagger}a_i-\frac{1}{2}).
\end{equation}
In the following it is understood that the energy eigenvalues have been enumerated in increasing order, so that $x_1\leq x_2\leq \hdots \leq x_{i-1}\leq x_i\leq x_{i+1}\leq \hdots\leq x_N$.  
We can of course consider any energy number $x_{F}$ obeying $x_{n_f}\leq x_{F}\leq x_{n_f+1}$ the fermisurface (energy), $n_f$ being the number fermions. Instead of keeping the number of fermions  fixed to $n_f$, we  include a chemical potential term in the \h ,
\begin{equation}
H_{\rm ch}=-x_{F}\cdot \# {\rm fermions}=-x_{F}\cdot n_f=-x_{F}\sum_i a_i^{\dagger}a_i.
\end{equation}  
With such a term one arranges the minimal energy situation to have precisely the (wanted) value $x_{F}$ for the Fermi surface.
Analogously to the fermion interaction term (\ref{pma}), we  shall also for this chemical potential term choose symmetrized expression $[a_i^{\dagger},a_i]$ in the annihilation and creation operators.
 If we consider the situation in which the fermions for a given state of the dynamical matrix $M$  had adjusted to lower the energy most possible (for fixed $M$) then  we would have the $n_f$ lowest eigenvalues $x_i$ filled and remaining $N-n_f$ eigenenergy levels empty. Including zero-point energy  the value of the $ H_{\rm int.\;fermion}$ part of the \h \ would be
\begin{equation}\label{intf}
\left.H_{\rm int.\;fermion}\right|_{\overset{\rm minimal}{\underset{\rm energy}{\tiny{\rm fermion}}}}=\sum_{i=1}^{n_f}\frac{1}{2}x_i-\sum_{i=n_f+1}^{N}\frac{1}{2}x_i \rightarrow - \frac{1}{2}\int dx \r(x) x {\rm sign}(x_F-x) , 
\end{equation}
and the chemical potential term would give
\begin{equation}
\left.H_{\rm ch}\right|_{\overset{\rm minimal}{\underset{\rm energy}{\tiny{\rm fermion}}}}=-\sum_{i=1}^{n_f}\frac{1}{2}x_F+\sum_{i=n_f+1}^{N}\frac{1}{2}x_F
\rightarrow  \frac{x_F}{2}\int dx \r(x) {\rm sign}(x_F-x) . 
\end{equation}

\section{Ground state and the spectrum distribution}

The full \h \ for the system would contain additional terms coming from fermion interaction and chemical potential terms 
\begin{equation}
H_{\rm full}=H+\left.H_{\rm int.\;fermion}\right|_{\overset{\rm minimal}{\underset{\rm energy}{\tiny{\rm fermion}}}} +\left.H_{\rm ch}\right|_{\overset{\rm minimal}{\underset{\rm energy}{\tiny{\rm fermion}}}}.
\end{equation}
Properly it should, however, be stressed that although we shall formally use this expression it is in fact only physically justified when one seeks the ground state. If the system gets excited  we also expect the fermion part of it to get excited and then one should in principle treat the fermionic part of the system as a properly second-quantized system. 

In seeking the minimum of effective potential of our dynamical model 
\begin{equation}\label{veff}
V_{\rm eff}=\frac{1}{2}\int dx\left\{ \frac{\pi^2}{3}\r^3(x)+\o^2x^2\r(x)-\mu \r(x)-(x_F-x){\rm sign}(x_F-x)\r(x)\right\},
\end{equation} 
and solving $\delta V_{\rm eff}/\delta\rho=0$  one obtains the  semiclassical ground-state eigenvalue density
\begin{equation}\label{rr}
\r(x)=\frac{1}{\pi}\sqrt{\mu+(x_F-x){\rm sign}(x_F-x)-\o^2x^2}.
\end{equation}
For $x<x_F$, i.e., for filled states we have
\begin{equation}
\r(x)=\frac{ \o}{\pi}\sqrt{\frac{\tilde{\mu}+x_F}{\o^2}-\left(x+\frac{1}{2\o^2}\right)^2},
\end{equation}
and for empty states, $x>x_F$, we have
\begin{equation}
\r(x)=\frac{ \o}{\pi}\sqrt{\frac{\tilde{\mu}-x_F}{\o^2}-\left(x-\frac{1}{2\o^2}\right)^2},
\end{equation}
Pictorially, we obtained semicircle law for both, filled and empty states, separated by the gap, when the chemical potential $\tilde{\mu}=\mu+(2\o)^{-2}$ satisfies $x_F<\tilde{\mu}<x_F+1/\o^2$, and $0<x_F<1$, see Fig.1.
\begin{figure}[h]
\begin{center}
\includegraphics*[width=8.0cm]{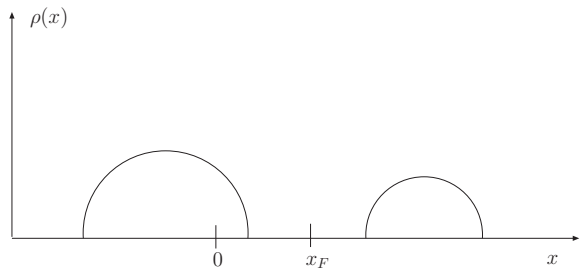}
\end{center}
 \caption{On the left we have filled levels and on the right empty ones.}
 \label{fig2}
\end{figure}
Notice that the term $(x-x_F){\rm sign}(x-x_F)$ in $V_{\rm eff}$ has a constant derivative, corresponding to a constant 'electric field', on each side of the Fermi surface $x_F$. This 'electric field' pulls the energy levels away from $x_F$, producing the homolumo gap. If we didn't symmetrized the interaction with respect to filled and empty states, we would have $\theta(x-x_F)$ instead of  ${\rm sign}(x-x_F)$, i.e. we would still obtain the gap, only it wouldn't be symmetric with respect to $x_F$.


A more intuitive physical picture  could be, however, obtained in the following way.
As we are interested only in the ground state we approximate the fermionic interaction with a smooth polynomial expression, as depicted in Fig.2. 
\begin{figure}[h]
\begin{center}
\includegraphics*[width=8.0cm]{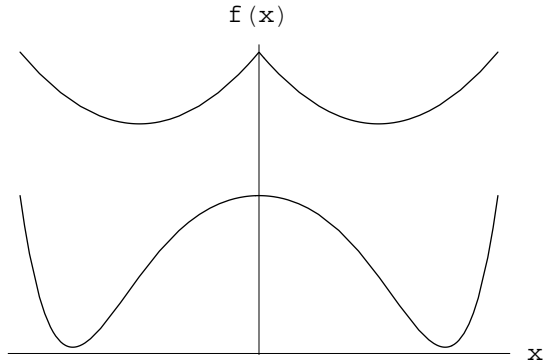}
\end{center}
 \caption{Smoothing the potential: upper graph is $f(x)=x^2-|x|$, and lower is $f(x)=-x^2-x^4+2x^6$.}
 \label{fig1}
\end{figure}
Choosing $f(x)=a x^6-b x^4- c x^2$ we obtain for the ground-state functional 
\begin{equation}
\Phi=\exp\left\{\frac{1}{2} \int dx dy \r(x)\ln (x-y)\r(y)-\frac{1}{2}\int dx \r(x)(fx^4-gx^2)\right\},
\end{equation}
This correspond to 
\begin{equation}\label{rmtquad}
Z=\int \prod_{i} dx_i \prod_{i< j}(x_i-x_j)^{2
} e^{-f\sum_i x_i^4+g\sum_i x_i^2},
\end{equation}
giving, for deep enough wells, the two-cut solution.
This shows that a homolumo gap can appear in a very generic system. Maybe even too generic, as it is basically just two-cut solution of a hermitean matrix model. However, the physical interpretation and field-theory formulation that we presented enables one to go beyond the ground state using rather standard techniques.    Remember that it has been shown \cite{mehta} that there are a few eigenvalues left outside of the semicircle, meaning that there are few states left in the gap. Thinking of our model as  a field-theoretical model one can look for the  instanton contributions \cite{aji} which are expected to give strongly-reduced but non-zero  level density near the Fermi surface.
Detail analysis of these effects is left for further investigation.

One purpose of the presented studies in a very general setting is the application of the homolumo gap effect in the project of Random Dynamics \cite{hf}. There, one starts from the observation that at the present the energies at our disposal are extremely low, compared to the supposed fundamental energy scale, presumably to be identified with the Planck scale. Thus the 'poor physicist' can only hope to bring say a fermion from just below the Fermi-energy to just above it from the fundamental scale point of view. Obviously, the appearance of a homolumo gap of the fundamental energy scale order of magnitude would prevent creation of any fermion whatsoever. What we are really interested in is obtaining strong reduction of level density near the Fermi surface, instead of a genuine homolumo gap. Namely, if the level density in the energy range we consider is especially low, then whenever a potential scattering from one momentum eigenstate to one with a different eigen-momentum should take place there will be for it especially small phase space available. That is to say there would be very few states to scatter into and thus we expect much fewer such momentum violating scatterings would take place. This strong enhancement in the accuracy of momentum conservation  would be interpreted as a Random Dynamics spirit derivation of momentum conservation.

\section*{Acknowledgments}
This work was supported by the Ministry of Science, Education and Sports of the
Republic of Croatia under the contract 098-0982930-2861.

\end{document}